\documentclass[11pt,a4paper]{article}
\pdfoutput=1

\usepackage[]{Packages}

\newcommand*\circled[1]{\tikz[baseline=(char.base)]{
   \node[shape=circle,draw,inner sep=1.2pt] (char) {#1};}}

\bibliography{GaugeBetheRef}

\usepackage{Definitions}

\preprint{\texttt{CERN-PH-TH/2013-232}}

\newcommand{\OfficialTitle}{A stringy perspective on the quantum integrable model/gauge correspondence}

\title{\vspace{2cm}
  {\huge   \textbf{\dosserif\OfficialTitle}}
}

\hypersetup{pdfauthor={Domenico Orlando},pdftitle={\OfficialTitle}}

\author{%
  \begin{minipage}{.8\linewidth}
    \vspace{1cm}
    \begin{center}
      {\small \textbf{Domenico Orlando} }
    \end{center}
    \vspace{1cm}
    \begin{minipage}{\linewidth}\centering
      {\itshape \footnotesize
        Theory Group, Physics Department, \\ Organisation européenne pour la recherche nucléaire (CERN) \\ CH-1211 Geneva 23, Switzerland
      }
    \end{minipage}
  \end{minipage}
}

\date{}

\begin{document}

\setstretch{1.1}

\numberwithin{equation}{section}

\begin{titlepage}

  \maketitle

  \thispagestyle{empty}

  \vfill
  \abstract
  {We present a string theory realization for the correspondence between quantum integrable models and supersymmetric gauge theories. The quantization results from summing the effects of fundamental strings winding around a compact direction. We discuss the examples of the \textsc{xxz} gauge/Bethe correspondence and five-dimensional \( \Omega \)--deformed \textsc{sym} on \( M \times S^1 \).}
\vfill

\end{titlepage}

\section{Introduction}
\label{sec-1}

  The deep relationship between supersymmetric gauge theories and integrable models has proven valuable for both disciplines since the seminal work of Donagi and Witten~\cite{Donagi:1995cf}. In the recent years this type of analysis relating (deformed) gauge theories to integrable models in various dimensions has found some remarkable realizations such as the \textsc{agt} correspondence~\cite{Alday:2009aq} or the \textsc{bps}/Bethe correspondence~\cite{Nekrasov:2009uh,Nekrasov:2009ui}. In the former, the partition function of \( \mathcal{N} = 2 \) \textsc{sym} on \( S^4 \) is related to the conformal blocks of Liouville field theory; in the latter the ground states of a two-dimensional \( \mathcal{N} = (2,2) \) system with twisted masses are identified with the Bethe states of an integrable \textsc{xxx} spin chain.
Both examples have a natural string/M--theory realization that can be used to understand the underlying structure implying the correspondence of the supersymmetric and integrable models~\cite{Orlando:2010uu,Orlando:2010aj,Hellerman:2011mv,Hellerman:2012zf,Hellerman:2012rd,Lambert:2013lxa,Tan:2013tq,Yagi:2012xa}.

From the integrable model point of view, most systems admit a natural extension corresponding to passing from classical to quantum symmetry. The archetypal example is the \textsc{xxx} spin chain with \( sl_2 \) symmetry and its \textsc{xxz} generalization with \( U_q(sl_2) \) symmetry\footnote{This system has also appeared in connection to supersymmetry in~\cite{Dijkgraaf:2008ua,Orlando:2009kd}.}. It is natural to wonder if such a quantization on the integrable side has a natural counterpart on the supersymmetric side. The answer seems to be affirmative as shown by some examples in the recent literature~\cite{Nekrasov:2009ui,Nieri:2013yra,Tan:2013xba,Itoyama:2013mca,Awata:2010yy}. We will show that all these examples follow a common pattern. The original gauge theory on a manifold \( M \) is deformed by adding a set of states that can be seen as the full Kaluza--Klein tower coming from the reduction of a  higher-dimensional system on \( M \times S^1 \) (or, more generally, an \( S^1 \)--fibration over \( M \)). The quantum parameter of the corresponding integrable model is proportional to \( q \propto e^R \), where \( R \) is the radius of the \( S^1 \).

The aim of this note is to introduce a string theoretical framework in which such theories can be realized in terms of D--branes in non-trivial spacetimes with fluxes including, in some frames, a constant \( B \)--field.  Such a field is typically an equivalent description of non-commutative geometries, which are related to the notion of quantum geometry associated to quantum groups. We will also lift these brane configurations to M--theory, where the extra Kaluza--Klein modes are interpreted as a gas of \M2--branes.

\section{Kaluza--Klein modes, B--field and twisted masses}
\label{sec-2}

The gauge theory that we intend to realize is the lift to \( \mathbb{R}^2 \times S^1 \) of a two-dimensional gauge theory on \( \mathbb{R}^2 \) in which all the Kaluza--Klein excitations are kept explicitly. Concretely, if \( R \) is the radius of the \( S^1 \), the two-dimensional theory includes an additional tower of states with mass
\begin{equation} 
  \label{eq:KK-modes}
  E_w = \frac{|w|}{R} \, , \hspace{2em} w \in \mathbb{Z} \,.
\end{equation}

Even though there is a natural interpretation of these states in terms of winding modes of the string theory around the T--dual circle, the most naive realization of the gauge theory in terms of a \textsc{dbi} action fails to take these states into account as it is limited to the massless degrees of freedom in the theory.

Let us consider more closely the three-dimensional picture. The Kaluza--Klein spectrum is reproduced if we break the invariance under translations on the \( S^1 \) by adding a corresponding twisted mass \( \tilde m = 1/R \).  The simplest way of writing such a theory~\cite{Hellerman:2011mv} is to consider the fibration \( S^1 \to \mathbb{R}^2 \times T^2 \to \mathbb{R}^2 \times S^1 \) defined by the monodromy
\begin{align} 
  x^2 \simeq x^2 + 2 \pi R\,, &&\begin{cases}
  x^8 \simeq x^8 + 2 \pi R_8\,, \\
  x^2 \simeq x^2 + 2 \pi R_8 R \tilde m\,,
\end{cases}
\end{align} 
where \( x^8 \) parametrizes the fiber of radius \( R_8 \) and \( x^2 \) parametrizes the \( S^1 \) in the base. In other words one should look at the lift of the three-dimensional theory to \( \mathbb{R}^2 \times T^2 \), where \( T^2 \) is the torus with parameter \( \tau = R_8 \left( \tilde m + \frac{i}{R} \right) \).

We have three equivalent descriptions of the same physical system:
\begin{enumerate}[label=\protect\circled{\arabic*}]
\item a two-dimensional gauge theory on \( \mathbb{R}^2 \) with an additional tower of modes of mass \( E_w = |w|/ R \)
\item a three-dimensional gauge theory on \( \mathbb{R}^2 \times S^1 \)  with twisted mass \( \tilde m = 1/R \) for the translations on \( S^1 \)
\item a four-dimensional gauge theory on \( \mathbb{R}^2 \times T^2 \) on a torus with parameter \( \tau = R_8/R \left( 1 + i \right) \), where \( R_8 \) is an auxiliary parameter that is consistently sent to zero in the other pictures.
\end{enumerate}

Our strategy is as follows. We first realize the theory in the third picture in terms of the effective action for a \D3--brane and then we work our way up to the other pictures using T--duality. The result will be an explicit form for the three-dimensional theory and a winding string interpretation for the extra modes of the two-dimensional theory.

Let us start with picture \circled{3}. In order to realize the \( \mathbb{R}^2 \times T^2 \) theory as a D--brane effective action, we must start with a bulk of the type \( \mathbb{R}^2 \times T^2 \), where the torus has the parameter \( \tau = R_8/R \left( 1 + i \right) \). Consider flat coordinates \( x^\mu, \mu = 0, \dots, 9 \) with the identifications
\begin{align} 
  x^2 \simeq x^2 + 2 \pi R\,, &&\begin{cases}
    x^8 \simeq x^8 + 2 \pi R_8\,, \\
    x^2 \simeq x^2 + 2 \pi R_8 \,.
  \end{cases}
\end{align} 
It is convenient to introduce a new \( (2 \pi R) \)--periodic coordinate \( y \),
\begin{equation}
  y = x^2 - x^8 \,,
\end{equation}  
such that the bulk metric takes the form of a fibration of \( x^8 \) over \( \mathbb{R}^8 \times S^1 \):
\begin{equation} 
  \di s^2 = \di \mathbf{x}_{01345679}^2 + \frac{1}{2} \di y^2 + 2 \left( \di x^8 + \frac{1}{2}\di y \right)^2.
\end{equation}  
The four-dimensional gauge theory \circled{3} is realized as the \textsc{dbi} action for a \D3--brane extended in \( (x^0, x^1, y, x^8) \) and its gauge coupling is \( g^2 = \pi e^{\Phi_0}  \), where \( \Phi_0 \) is the constant value of the bulk dilaton.
The advantage of this coordinate system is that the periodicities of \( x^8 \) and \( y \) are independent and we can decouple the (unphysical) modes around \( x^8 \) by T--dualizing and taking the \( R_8 \to 0 \) limit.

\bigskip 

After T--duality in \( x^8 \) we move to picture \circled{2} where the connection of the \( S^1 \)--fibration has turned into a B--field. The bulk fields are given by
\begin{subequations}
  \begin{align}
    \di s^2 &= \di  \mathbf{x}_{01345679}^2 + \frac{1}{2} \di y^2 + \frac{1}{2} (\di  \tilde x^8)^2 \,,\\
    B &= \frac{1}{2} \di y \wedge \di  \tilde x^8\,, \\
    e^{-\Phi} &= \sqrt{2} e^{-\Phi_0}\,.
  \end{align}
\end{subequations}
A few remarks are in order.
\begin{enumerate}
\item The metric still has the form \( \mathbb{R}^8 \times T^2 \), but now the torus is rectangular, with independent identifications \( y \simeq y + 2 \pi R \) and \( \tilde x^8 \simeq \tilde x^8 + 2 \pi \alpha'/R_8 \);
\item a constant \( B \)--field has appeared. Its field strength vanishes and in this it differs from the field that appears in the (related) fluxtrap description of the \( \Omega \)--deformation. The fundamental difference is that in the case at hand we introduce twisted masses to mod out a \( U(1) \) symmetry that acts \emph{freely} on the geometry;
\item no supersymmetries are broken by the identifications and T--duality, once more because we are considering a freely-acting orbifold.
\end{enumerate}
After T--duality, the \D3--brane turns into a \D2--brane extended in \((x^0, x^1, y)  \). The corresponding \textsc{dbi} action in static gauge is similar to the one of the \( \Omega \)--deformation. The relevant part of the bosonic component reads
\begin{equation}
  \mathcal{L} = \frac{1}{g^2} F^{\mu \nu} F_{\mu \nu} + \left( \partial_\mu \sigma_1 + \frac{1}{g} V^\nu F_{\nu \mu} \right)^2 + \dots \,,
\end{equation}
where \( V \) is the unit vector in the direction \( y \),
\begin{equation} 
  V^\nu \partial_\nu = \partial_y \,,
\end{equation}  
and \( \sigma_1 \) is the field that describes the motion of the brane in the direction \( \tilde x^8 \), which we are free to consider non-periodic after having taken the \( R_8 \to 0 \) limit (recall that \( \tilde x^8 \) has radius \( \alpha'/R_8 \)).
If  the potential \( A_\mu \) does not depend on \( y \), the three-dimensional action is obtained by promoting the field \( \sigma_1 \) to a covariant derivative ,
\begin{equation} \sigma_1 \mapsto \sigma_1 + \frac{1}{g} A_y \,,\end{equation}
as it was already discussed in~\cite{Nekrasov:2009uh}.
What we find is another manifestation of the effect of a closed bulk \( B \)--field on the dynamics of a D--brane. In fact, this construction in which the brane is wrapped around a circle (the direction \( y \)) in the torus \( (y, \tilde x^8) \) that supports the \( B \)--field is in some sense intermediate between the situation studied by Seiberg and Witten~\cite{Seiberg:1999vs} in which the brane covers the full torus and the one studied by Douglas and Hull~\cite{Douglas:1997fm} where the brane is a point in the torus. In both cases the effective theories admit a natural interpretation in terms of non-commutative geometry.

\bigskip

Finally, we can move to picture \circled{1} by T--dualizing in \( y \). The \( B \)--field vanishes, the dilaton takes the same value \( \Phi_0 \) as in frame \circled
{3} and the metric takes the form
\begin{equation} 
  \di s^2 = \di \mathbf{x}_{01345679}^2 + \left( \di \tilde x^8 - \di \tilde y \right)^2 + \di \tilde y^2 \,,
\end{equation}
where \( \tilde y \) is periodic with period \( 2 \pi \alpha'/R \). The \D2--brane turns into a \D1--brane extended in \( (x^0, x^1) \). The original Kaluza--Klein modes of the gauge theory can now be understood as modes of the fundamental strings wound around \( \tilde y \). The energy of a string winding \( w \) times is then
\begin{equation} E_w = T \ell = \frac{1}{2 \pi \alpha'} 2 \pi \frac{\alpha'}{R} w = \frac{w}{R}\end{equation}
in perfect agreement with Equation~\eqref{eq:KK-modes}.

\section{Applications}
\label{sec-3}

The construction that we have outlined so far can be combined with other deformations of gauge theories in various dimensions to obtain a string realization of examples of interest that have appeared in the literature in connection with quantum integrable models.

In~\cite{Nekrasov:2009uh} the authors study a \( U(N) \) \( \mathcal{N} = 2^* \) two-dimensional gauge theory with twisted mass for the adjoint and a tower of Kaluza--Klein modes. They find that the supersymmetric ground states for the system are described by the Bethe Ansatz equations for an \textsc{xxz} spin chain with \( U_q (sl_2) \) symmetry with
\begin{equation} q = e^{- m R}\,, \end{equation}
where \( m \) is the twisted mass of the adjoint field and \( R \) is the radius of the Kaluza--Klein reduction. The \textsc{xxx} theory (without Kaluza--Klein modes) was already realized in~\cite{Hellerman:2011mv} in terms of \D2--branes suspended between \NS5--branes in the fluxtrap background. A similar construction can be repeated here. The equivalent of picture \circled{3} is obtained by starting with a \NS5--\D4 system with \NS5--branes extended in \( x^0, \dots, x^3, x^8, x^9 \) and a \D4 suspended between the \NS5s and extended in \( x^0, x^1, x^6, x^8, x^9\) with identifications
\begin{align} 
x^9 \simeq x^9 + 2 \pi R\,, &&\begin{cases}
  x^8 \simeq x^8 + 2 \pi R_8\,, \\
  x^9 \simeq x^9 + 2 \pi R_8\,, \\
  \theta_2 \simeq \theta_2 + 2 \pi R_8 m\,, \\
  \theta_3 \simeq \theta_3 - 2 \pi R_8 m  \,, 
\end{cases}
\end{align} 
where \( \tan \theta_2 = x^3/x^2 \) and \( \tan \theta_3 = x^5/x^4 \). Following~\cite{Hellerman:2011mv} we disentangle the periodicities by introducing the variables
\begin{align} 
  y &= x^9 - x^8 \,, &
  \phi_2 &= \theta_2 - m x^8\,, &
  \phi_3 &= \theta_3 + m x^8 \,.
\end{align}
After a T--duality in \( x^8 \) the \D4 turns into a \D3 and the effective description is a three-dimensional theory on \( \mathbb{R}^2 \times S^1 \) with real mass \( m \)  for the adjoints and \( 1/R \) for the translations in the \( S^1 \) (picture \circled{2}).
A final T--duality in \( y \) turns the brane configuration into an \NS5--\D2 system with extra modes corresponding to windings of the fundamental string in \( \tilde y \) (picture \circled{1}).

\bigskip

In~\cite{Nekrasov:2002qd,Nekrasov:2003rj} the authors evaluate the partition function for a \( U(1) \) five-dimensional non-commutative theory on \( \mathbb{R}^4 \times S^1 \) in the \( \Omega \)--background and they find that it can be understood as a \( q \)--deformation of the standard \( d=4 \) \( \mathcal{N} =2  \) theory with two \( \epsilon \) parameters, where \( q_i = exp[ \epsilon_i  R] \). By now we can recognize this as the higher-dimensional analogue of the previous construction, \emph{i.e.} as the effective theory in picture \circled{2} of a \D5--brane suspended between two \NS5--branes. 
In picture \circled{3} this corresponds to a \D6--brane extended in \( (x^0, \dots, x^3, x^6, x^8, x^9) \) between two \NS5--branes in a Melvin spacetime with identifications~\cite{Hellerman:2012zf}
\begin{align} 
  x^9 \simeq x^9 + 2 \pi R\,, && \begin{cases}
    x^8 \simeq x^8 + 2 \pi R_8\,, \\
    x^9 \simeq x^9 + 2 \pi R_8\,, \\
    \theta_1 \simeq \theta_1 + 2 \pi R_8 \epsilon_1\,, \\
    \theta_2 \simeq \theta_2 + 2 \pi R_8 \epsilon_2\,, \\
    \theta_3 \simeq \theta_3 + 2 \pi R_8 \epsilon_3\,,
  \end{cases}
\end{align} 
where \( \tan \theta_1 = x^1/x^0 \) and in order to preserve supersymmetry, \( \epsilon_1 + \epsilon_2 + \epsilon_3 = 0\).
After introducing proper angular coordinates to disentangle the periodicities we can T--dualize in \( x^8 \) to go to picture \circled{2}, where the \D5--brane describes the five-dimensional picture. 
A T--duality in the \( S^1 \)--direction turns the system into picture \circled{1} with a \D4--brane suspended between \NS5s with fundamental strings wrapping the dual circle. This latter system can be lifted directly to M--theory. If we consider the case in which \( x^6 \) is periodic, the \D4--\NS5 system lifts to an \M5--brane extended on \( \mathbb{R}_{\epsilon_1, \epsilon_2} \) and wrapped on a torus with a puncture. The fundamental strings are lifted to \M2--branes wrapped on the torus and on an extra circle of radius proportional to \( 1/R \) as shown in Table~\ref{tab:brane-diagram}.

An equivalent description for this gas of \M2--branes can be obtained by using the \( SL_2 (\mathbb{Z}) \times SL_3(\mathbb{Z}) \) symmetry of M--theory on \( T^3 \) generated by \( \langle x^6, \tilde x^8, x^{10} \rangle \) that leaves the \M5--brane invariant and turns the \M2--branes wrapped on two cycles of the \( T^3 \) into Kaluza--Klein excitations~\cite{Polchinski:1998rr}\footnote{The reduction of this gas of Kaluza--Klein modes leads to a \D4/\D0 system in type IIA. This realizes the theory as described in~\cite{Nekrasov:1996cz,Lawrence:1997jr}.} along the third circle \( \tilde x^8 \).

In both descriptions we get the picture of the theory of an \M5--brane embedded into a non-trivial spacetime, which \emph{cannot be described by a conventional six-dimensional gauge theory}.

\begin{table}
\centering
\begin{tabular}{lccccccccccc}
    \toprule
    object   & $x_0$  & $x_1$       & $x_2$  & $  x_3$         & $x_4$ & $x_5$ & $x_6$      & $x_7$     & $\tilde x_8$     & $\tilde y$     & $x_{10}$       \\ 
\midrule
    M5      & $\times$  & $\times$       & $\times$  & $\times$           &          &        & $\times$   &           &           &    &$\times$         \\      
    M2      &   &       &   &            &          &        & $\times$   &           &           &     $\times$       & $\times$ \\      
\bottomrule
\end{tabular}
\caption{Brane diagram for the M5--M2 system that realizes the five-dimensional gauge theory on \( \mathbb{R}^4_{\epsilon_1, \epsilon_2} \times S^1 \).}
\label{tab:brane-diagram}
\end{table}

\bigskip

In~\cite{Nieri:2013yra} the authors describe the correspondence between the partition function on \( S^4 \times S^1 \) and conformal blocks of quantum Liouville field theory with parameter \( q = e^{\epsilon R} \). This system can be seen as the picture \circled{2} for a system analogous to the previous one, where the M5--brane is now wrapped on \( S^4_{\epsilon_1, \epsilon_2} \times T^2 \) and the Kaluza--Klein modes running on the \( S^1 \) correspond to a gas of M2--branes wrapping the torus and an external compact direction of radius proportional to \( 1/R \).
A similar description is possible for the \( S^5 \) case: now we need to consider the fibration \( S^1 \to S^5 \to \mathbb{P}^2 \) and it is natural to conjecture that the system will be realized as an M5--brane wrapped on \( \mathbb{P}^2_{\epsilon_1, \epsilon_2} \times T^2 \) with a gas of M2--branes. An explicit realization of these systems is currently under investigation.  

\section*{Acknowledgements}

I would like thank Sara Pasquetti and Filippo Passerini for illuminating discussions and Susanne Reffert for insightful comments on the manuscript. Moreover I acknowledge the Galileo Galilei Institute for Theoretical Physics for hospitality, and the \textsc{infn} for partial support during the completion of this work.

\printbibliography

\end{document}